\documentclass[aps,prl,preprint,groupedaddress,amsmath,showkeys,showpacs]{revtex4}

\usepackage{graphicx}
\usepackage{dcolumn}
\usepackage{bm}

\begin{document}


\title{The superferromagnetic state in the ensemble of oriented Stoner-Wohlfarth particles: a coercivity due to phase stability}



\author{A.A. Timopheev}
\email[e-mail: ]{ timopheev@iop.kiev.ua}
\affiliation{Institute of Physics NAS of Ukraine, Prospect Nauki str. 46, Kiev, 03028, Ukraine}

\author{V.M. Kalita}
\affiliation{Institute of Physics NAS of Ukraine, Prospect Nauki str. 46, Kiev, 03028, Ukraine}

\author{S.M. Ryabchenko}
\affiliation{Institute of Physics NAS of Ukraine, Prospect Nauki str. 46, Kiev, 03028, Ukraine}

\author{A.F. Lozenko}
\affiliation{Institute of Physics NAS of Ukraine, Prospect Nauki str. 46, Kiev, 03028, Ukraine}

\author{P.A. Trotsenko}
\affiliation{Institute of Physics NAS of Ukraine, Prospect Nauki str. 46, Kiev, 03028, Ukraine}

\author{M. Munakata}
\affiliation{Energy Electronics Laboratory, Sojo University, Kumamoto 860-0082, Japan}


\date{\today}

\begin{abstract} 

It is observed experimentally that the coercive field has an anomalous angular dependence at temperatures above the blocking temperature in physically nonpercolated granular films CoFeB-SiO$_{2}$ with anisotropic granules oriented in the same direction. It is shown that the anomaly is determined by the singularity of an angular dependence of the critical field causing the absolute loss of phase stability of the superferromagnetic state of an ensemble of interacting superparamagnetic granules.

\end{abstract}



\maketitle

\section{\label{body}}

Ensembles of single-domain ferromagnetic nanoparticles are of interest from the applied and theoretical viewpoints. The perspective to use the ensembles of oriented Stoner-Wohlfarth particles \cite{ref17} (SW-particles) as high-density recording media stimulates experimental studies of such objects. The most significant factors restricting their use in this field are the ``superparamagnetic limit'' and the interparticle interaction. Therefore, the clarification of the role of the mentioned factors attracts a special attention of researchers. An intergranular interaction will lead to the space-time correlation of magnetic moments of granules in an ensemble. If such an interaction of the ferromagnetic type, a particular state -- the superferromagnetic (SF) one -- of an interacting superparamagnetic ensemble with a characteristic ordering temperature $T_{sf}$ can arise. Papers  \cite{ref1,ref2,ref3,ref4,ref5,ref6,ref7,ref8,ref9,ref10,ref11,ref12} present the experimental data obtained within various methods for ensembles of different types which indicate the appearance of the SF state. However, the nature of an interaction inducing the SF ordering of an ensemble and the stability of this state remains controversial. The reasons for SF ordering are considered to be the interparticle dipole-dipole interaction \cite{ref4,ref8,ref13}, various types of the exchange interaction \cite{ref11,ref14}, or their combination \cite{ref10}. Not completely clear are the questions concerning the joint manifestation of the SF state and characteristic peculiarities of superparamagnetic properties of ensembles such as the blocking, temperature dependence of a coercive field, \textit{etc}.

One may assume that properties of the SF phase of an ensemble of oriented SW-particles should be similar to properties of a classical anisotropic ferromagnet. However, the basic difference consists in the different relations between the effective interaction field $H_{int}$, the magnetic moment $J$ of a particle, and the anisotropy field $H_{a}$.

It is well known that the magnetization reversal of an anisotropic homogeneous (single-domain) ferromagnet is accompanied by a hysteresis. In a classical ferromagnet, the effective field of the inter-ion exchange interaction is much greater than $H_{a}$. Therefore, the critical field causing the magnetization reversal for a homogeneous ferromagnet is determined only by the anisotropy field. In a superparamagnetic ensemble, the interparticle interaction field $H_{int}$ can be less than $H_{a}$ (this case is often realized under a sufficient ``dilution'' of magnetic particles in a nonmagnetic matrix). We will show that the critical field causing the reversal of the magnetic moment of an ensemble in the SF phase will be determined in this case by the interaction, rather than the anisotropy. To the best of our knowledge, no studies of the effect of the relationship of the anisotropy of particles and their interaction on the magnetization reversal of the SF phase of an ensemble and the coercivity have been carried out earlier.

In a superparamagnetic ensemble of particles, the weak field $H_{int}$ of the interparticle interaction acts on the huge magnetic moment $J$ of a particle. This leads to that the SF ordering temperature $T_{sf}$ can be comparable with the blocking temperature $T_{b}$ of the ensemble. Depending on the interaction strength and the measurement duration (which can be different for different techniques), two cases are possible: \textit{i}) $T_{sf}<T_{b}$, and the magnetic properties (including the coercivity) of the ensemble will be mainly determined by a metastability of a blocked state; \textit{ii}) $T_{sf}>T_{b}$. In the latter case in the region $T_{b}<T<T_{sf}$, the properties of the SF phase become determining, and the magnetization reversal of the ensemble will have a nonrelaxation character. We will demonstrate that, in such case, the magnetization reversal of the uniformly magnetized SF phase in a magnetic field is an orientational first-order magnetic field-induced phase transition.

In the present paper, we study the magnetic properties of physically nonpercolated nanogranular CoFeB-SiO$_{2}$ films with anisotropic granules oriented in the same direction in the plane of a film. It will be shown that the conditions $H_{int}<H_{a}$ and $T_{sf}>T_{b}$ are satisfied in the film under study, and the coercivity of the film has anomalous angular dependence in the interval of temperatures $T_{b}<T<T_{sf}$. We will show that the anomalous angular dependence of the coercitivity in this interval is determined by a singularity of the angular dependence of the critical field causing the absolute instability of the SF phase. At temperatures below $T_{b}$, properties of the SF phase will be combined with blocking effects.

\begin{figure}
\begin{center}
\includegraphics[width=8.5cm]{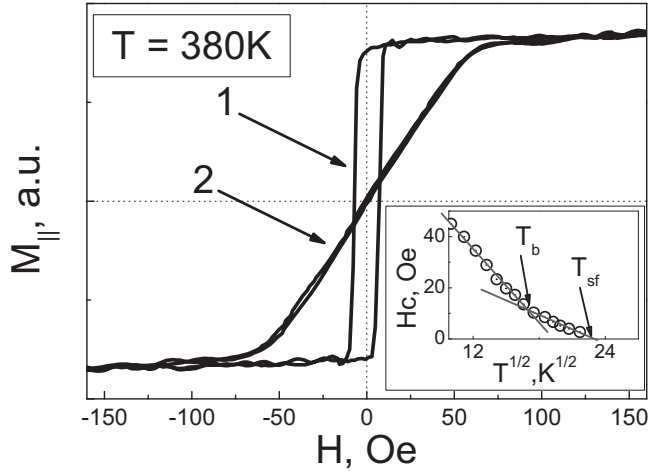}
\end{center}
\caption{\label{fig1} Curves of magnetization reversal along the easy and hard directions in the plane of a film at $T=380$ K. The insert shows the coercive field under the magnetization along the easy axis versus $T^{1/2}$.}
\end{figure}

Granular films (Co$_{0.25}$Fe$_{0.66}$B$_{0.09}$)$_{0.6}$-(SiO$_{2}$)$_{0.4}$ about 500 nm in thickness were grown by magnetron sputtering \cite{ref15,ref16}. The X-ray diffraction studies \cite{ref16} showed that the granules are amorphous. A film reveals the easy-plane anisotropy related to the demagnetization factor of the film as a whole ($4\pi M_{eff}\cong 9500$ Gs). The magnetization reversal curves (see Fig.\ref{fig1}) indicate that the film has also the intraplane anisotropy ($H_{a}\cong65$ Oe) which was induced due to a special growth technology and is, probably, related to a small ellipticity of granules \cite{ref15}. The measurements of magnetization reversal curves of the film (measurements of the projection of the film magnetization onto the direction of the applied field $M_{||}(H)$) were carried out with the help of a VSM-magnetometer LDJ-9500.

The magnetization reversal curves (see Fig.\ref{fig1}) of a specimen at $T=380$ K for the magnetization directions in the plane of the film at $\varphi=0^\circ$ and $\varphi=90^\circ$ ($\varphi$ is the angle between the easy axis of magnetization of the film and the direction of a magnetic field $H$) characterize unambiguously this specimen as an ensemble of oriented SW-particles \cite{ref17}. As the temperature decreases, curve 2 is practically invariable, and the rectangular form of curve 1 is conserved, but the coercivity ($H_{C}$) for curve 1 increases. The dependence $H_{C}(\sqrt{T})$ demonstrates the presence of two regions (see the insert in Fig.\ref{fig1}). The low-temperature region obeys the N\'{e}el-Brown law \cite{ref18} modified with regard for the interaction \cite{ref19}, and the high-temperature one describes a change of the coercivity of the SF state with variation of the temperature. According to the analysis performed in \cite{ref19}, the high-temperature and low-temperature regions correspond to the intervals $T_{b}<T<T_{sf}$ and $T<T_{b}$ respectively. In this case, the blocking temperature for this specimen $T_{b}\approx 350$ K, and the temperature of SF ordering $T_{sf}\approx550$ K. As seen from Fig.\ref{fig1}, even for $T>T_{b}$, the curve of magnetization reversal along the easy axis ($\varphi=0^\circ$) preserves the rectangular form with the almost $100\%$ remanence $M_{r}$. Since $M_{r}$ at $T_{b}<T<T_{sf}$ is determined by a spontaneous magnetization of the SF phase, we may assert that the SF state preserves a homogeneous ferromagnetic order during the magnetization reversal, and no SF domains similar to described in \cite{ref7,ref10,ref12} appear in the specimen. At the same time, no homogeneous magnetization of the film appears in the SF phase at a heating of the specimen to $T>T_{sf}$ and a subsequent cooling in $H=0$ (zero-field cooling), i.e. the mean magnetization value over the specimen is equal to zero in ZFC-mode.

\begin{figure}
\begin{center}
\includegraphics[width=16.5cm]{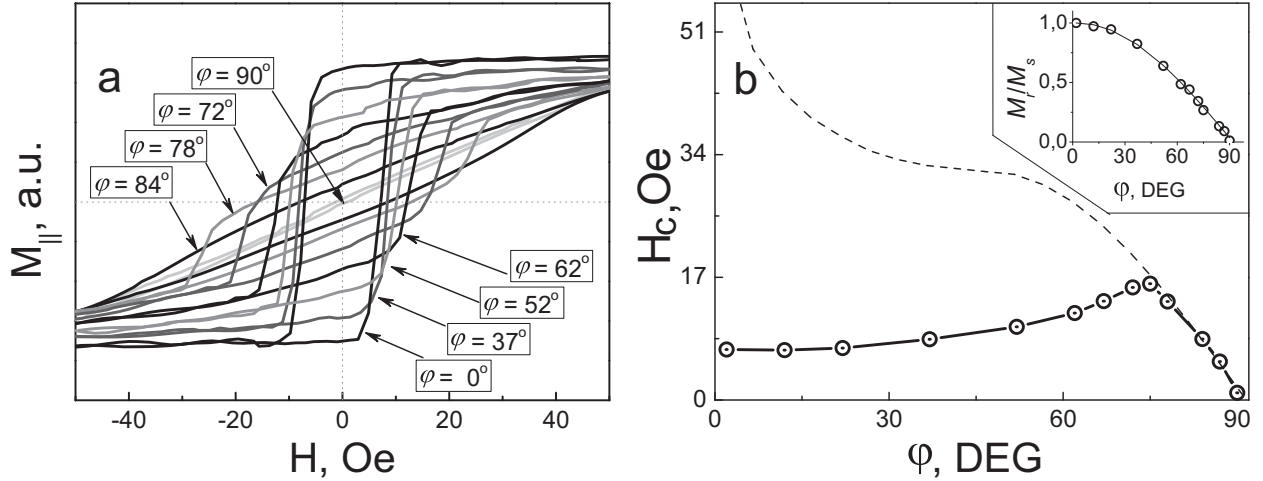}
\end{center}
\caption{\label{fig2} a) Magnetization reversal curves at $T=380$ K. b) Dependence $H_{C}(\varphi)$ at $T=380$ K (circles). The dotted line shows the expected dependence $H_{C}(\varphi)$ for an ensemble of noninteracting oriented SW-particles. The insert gives the angular dependence of the remanence of a specimen at the same temperature.}
\end{figure}

Figure \ref{fig2}a presents the dependences $M_{||}(H)$ obtained at $T=380$ K for various $\varphi$. As the angle $\varphi$ increases, the hysteresis loops become wider. At the same time, the remanence of the film ($M_{r}$) normed by its saturation magnetization ($M_{s}$) follows the dependence $M_{r}/M_{s}=\cos(\varphi)$. This means that the magnetic homogeneity of the SF phase (after the action of a saturating magnetic field) is preserved also in the process of magnetization reversal at arbitrary angles. The angular dependence of the coercivity $H_{C}(\varphi)$ shown in Fig.\ref{fig2}b demonstrates the presence of a maximum at $\varphi=\varphi_{\max}\cong65^{\circ}$ with $H_{C}(\varphi=\varphi_{\max})/H_{C}(\varphi=0^{\circ})\cong 2.5$. Such a behavior of $H_{C}(\varphi)$ is fundamentally different from the dependence of such a kind for an ensemble of noninteracting oriented Stoner-Wohlfarth particles which is shown in Fig.\ref{fig2}b by a dotted line (it has a maximum at $\varphi=0^{\circ}$ and demonstrates a decreasing angular dependence $H_{C}(\varphi)$ [17]).

\begin{figure}
\begin{center}
\includegraphics[width=8.5cm]{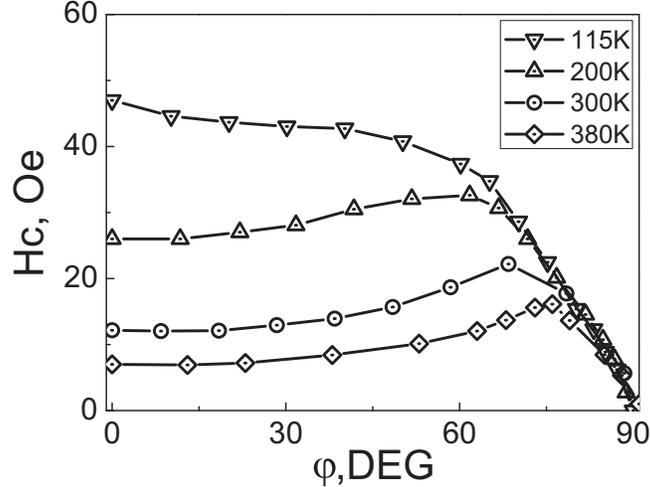}
\end{center}
\caption{\label{fig3} Modification of the dependence $H_{C}(\varphi)$ at the lowering of the temperature.}
\end{figure}

It is known \cite{ref20,ref21,ref23} that the angular dependence $H_{C}(\varphi)$ can have a maximum with $\varphi _{\max }\ne0$ even for noninteracting anisotropic particles at their magnetization reversal through the formation of an inhomogeneous magnetic state of particles (the so-called ``curling''). However, if namely this mechanism would be realized in our case, then the coercivity of a specimen at $\varphi =0$ would stop to increase with decrease in the temperature, being limited by conditions of the appearance of the curling in particles and, due to this fact, deviating from the N\'{e}el-Brown law. As seen from the insert in Fig.\ref{fig1}, the dependence $H_{C}(\sqrt{T})$ is linear for $T<T_{b}$. That is, the given law is fulfilled. As was established, the appearance of the maximum of $H_{C}(\varphi)$ with $\varphi_{\max}\ne0$ in our specimen is a temperature-dependent effect. It is dies out with decrease in the temperature below $T_{b}$ due to overlapping by the increasing coercivity of the blocked state (see Fig.\ref{fig3}). This confirms additionally that each granule in the process of magnetization reversal preserves its magnetic homogeneity, and the appearance of the maximum of $H_{C}(\varphi)$ at $\varphi_{\max }\ne0$ is related, in the case under consideration, to peculiarities of the magnetization reversal of the SF state of the ensemble.

We will show further that the obtained experimental result can be described by involving a homogeneous ordering of magnetic moments of granules in the mean-field approximation \cite{ref4,ref11,ref19} without detailing the nature of the interaction responsible for the ordering. We note that it is more natural to apply this approximation to the systems with an intergranular interaction of the exchange nature, though it can be applied to the description of the SF phase arising due to the dipole-dipole interaction as well \cite{ref13}.

Since the easy-plane anisotropy related to the demagnetization factor of a film ($4\pi M_{eff}\cong9500$ Gs) exceeds significantly the intraplane anisotropy of granules in the film ($H_{a}\cong65$ Oe), and $H$ lies in the plane of the film, we may assume that the magnetization vector also lies in the film plane during the magnetization reversal. Then the energy of a SW-particle can be written as

\begin{equation} \label{EQ1}
E=V_{p}(-K\cos^{2}(\varphi-\varphi _{m})-H\; m_{p}\cos(\varphi _{m})-\lambda\; m_{p}(m_{||}\cos(\varphi _{m}) +m_{\bot }\sin(\varphi_{m}))),
\end{equation}

where $K$ -- the anisotropy constant of a particle, $m_{p}$ -- its magnetization, $V_{p}$ -- its volume, $\varphi_{m}$ -- the angle between the magnetization of a particle and \textbf{H} in the plane, $\varphi$ -- the angle between the easy axis and \textbf{H}, \textbf{m} -- the mean (over the ensemble with regard for the filling factor) magnetization of a particle represented in Eq.(\ref{EQ1}) by two components: $m_{||}$ -- the component of \textbf{m} along \textbf{H}, and $m_{\bot}$ -- the component of \textbf{m} perpendicular to \textbf{H}, and $\lambda$ -- the coefficient of proportionality between \textbf{m} and the self-consistent field.

The calculation of the field dependence of the mean in-plane magnetization of the ensemble of such particles can be carried out with the help of the self-consistency equation for the components of \textbf{m}:

\begin{equation} \label{EQ2}
m_{||}=\frac{m_{p}}{Z}\int_{0}^{2\pi}\cos(\varphi_{m})\;exp(-E/(kT))d\varphi _{m}
\end{equation}

\begin{equation} \label{EQ3}
m_{\bot}=\frac{m_{p}}{Z}\int_{0}^{2\pi}\sin(\varphi_{m})\;exp(-E/(kT))d\varphi _{m}
\end{equation}

where $Z=\frac{m_{p}}{Z}\int_{0}^{2\pi}exp(-E/(kT))d\varphi _{m}$ -- the partition function, and $k$ -- the Boltzmann constant.

In the isotropic case (in the absence of easy-plane anisotropy) at $K\rightarrow0$ and $H\rightarrow0$, the equations for the mean field yield $T_{sf}=\lambda M^{2}/(3k)$ \cite{ref4}. In the case of a strong easy-plane anisotropy and at $K\rightarrow0$, relations (\ref{EQ1}-\ref{EQ3}) imply that the temperature of appearance of the SF state (with a spontaneous $M\ne0$ at $H\to0$) is $T_{sf}=\lambda M^{2}/(2k)$. For $K>>\lambda$, the temperature of appearance of the SF state is $T_{sf}=\lambda M^{2}/k$. For finite $K$ and a strong easy-plane anisotropy, $T_{SF}$ will lie between these two values.

Let us assume that the measurement time is so great, that the blocking effects can be neglected (by this, we realize the case where $T_{sf}>T_{b}$, like that in the experiment). We now model the curves of magnetization reversal in the interval of temperatures $T_{b}<T<T_{sf}$, by using Eqs.(\ref{EQ1}-\ref{EQ3}). It is convenient to reduce the parameters to the dimensionless form: $h=H/H_{a}$, where $H_{a}=2K/m_{p}$ -- the anisotropy field of a SW-particle (we assume that, for a SW particle, $m_{p}=const$), $T_{red}=kT/(K V_{p})$, $\lambda_{red}=\lambda m_{p}^{2}/(2K)$, $m_{red}^{||}=m_{||}/m_{p}$, $m_{red}^{\bot}=m_{\bot }/m_{p}$, and $h_{C}(\varphi)=H_{C}(\varphi)/H_{a}$.

\begin{figure}
\begin{center}
\includegraphics[width=16.5cm]{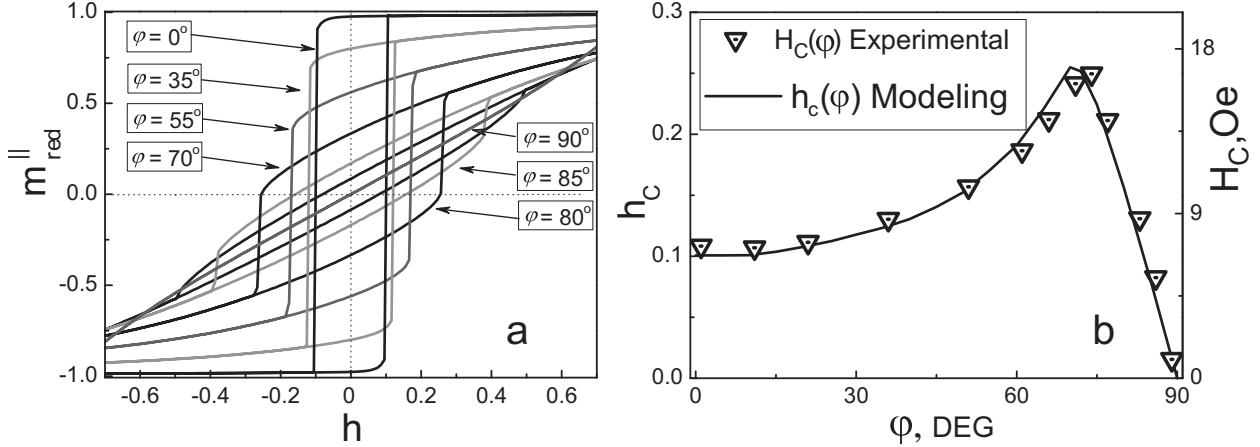}
\end{center}
\caption{\label{fig4} a) Model curves of magnetization reversal obtained at $T_{red}=0.1$, $\lambda_{red}=0.1$. b) Dependence $h_{C}(\varphi)$ obtained with these values of the parameters (solid line). Triangles show the data of $H_{C}(\varphi)$ from Fig.\ref{fig2}b.}
\end{figure}

In Fig.\ref{fig4}, we display the curves $m_{red}^{||}(h)$ and $h_{C}(\varphi)$ calculated with the help of the numerical solution of Eqs.(\ref{EQ1}-\ref{EQ3}) at $\lambda_{red}=0.1$ and $T_{red} =0.1$. Their comparison with the curves $M_{||}(H)$ and $H_{C}(\varphi)$ (Fig.\ref{fig2}) demonstrates a good qualitative agreement.

At phase transitions of the first order driven by a magnetic field, the self-consistency equations (\ref{EQ1}-\ref{EQ3}) allow us to obtain the course of the dependences $m_{red}^{||}(h)$ and $m_{red}^{\bot}(h)$ up to the instability points of the SF phase, i.e., in the region of metastability, where the external magnetic field is directed against $m_{red}^{||}(h)$. In the equilibrium thermodynamics, the point of a phase transition of the first order is determined from the condition of equality of the free energies. For the first-order phase transition driven by a magnetic field, it should correspond to the field $h=0$ in our model. However, the uniaxial anisotropy holds a metastable SF state. If the interparticle interaction dominates over the anisotropy, then the stability loss will be determined by the field $H_{a}$. In the opposite case realized in the granular film under study, a metastable SF state will be preserved up to the field causing its absolute instability. Such a field is determined by the interparticle interaction. The solutions of Eqs.(\ref{EQ1}-\ref{EQ3}) confirm this pattern. Moreover, according to these solutions, the angular dependence of the coercive field has the form corresponding to that observed in experiments. The experimental results indicate that the metastability of the SF phase is long-lived. This is related to the presence of the anisotropy of particles which is oriented in the same direction and stabilizes the total moment of the ensemble along a direction corresponding to the metastable energy minimum separated by a energy barrier from the global minimum.

The calculations performed on the basis of Eqs.(\ref{EQ1}-\ref{EQ3}) show that if the interaction field becomes stronger than the anisotropy field of a particle, i.e., $\lambda_{red}>1$, then the stability of the SF phase will be already determined by the anisotropy and the dependence $h_{C}(\varphi)$ will be determined by a Stoner-Wohlfarth astroid.

It is worth noting that $h_{C}$ coincides with the stability boundary of the SF state only in the angular interval $0<\varphi<\varphi_{\max}$. At $\varphi_{\max}<\varphi<\pi/2$ and $h=h_{C}(\varphi )$, the mean magnetic moment turns out to be turned normally to the applied magnetic field, and the magnetization jump (loss of the SF phase stability) occurs at $h>h_{C}$. The position of $\varphi_{\max}$ depends on the temperature, the interaction parameter, and the anisotropy of SW-particles. The detailed discussion of these dependences will be given elsewhere.

We note that similar maxima of the dependence $H_{C}(\varphi)$ for anisotropic particles with $0<\varphi_{\max}<\pi/2$ were observed earlier in other systems \cite{ref20,ref21,ref22,ref23}. They were explained exclusively by the magnetization reversal of a separate particle through its magnetically inhomogeneous state (curling). In the present work, we have shown that such an anomaly appears also at the preservation of the magnetic homogeneity of anisotropic particles in the case where the SF state arises in an ensemble with the anisotropy field exceeding the interparticle interaction field.

Thus, we have demonstrated that the angular dependence of a considerable coercivity, which is observed at $T>T_{b}$ in a physically nonpercolated granular film (CoFeB)$_{0.6}$-(SiO$_{2}$)$_{0.4}$ with anisotropic identically oriented single-domain granules, has a maximum for the angles close to the hard direction in the film plane. Such a type of the $H_{C}(\varphi)$ dependence does not correspond fundamentally to the Stoner-Wohlfarth model (and to the N\'{e}el model at $T>0$). Within a simple model, it is shown that such an anomaly of the angular dependence of the coercivity is related to the existence of the SF state at $T>T_{b}$ in the studied film under conditions where the anisotropy field of particles is stronger than the interparticle interaction field. In this state, the magnetic anisotropy of particles holds a metastable (with the magnetization directed against the magnetic field) state of the SF phase during the magnetization reversal up to the point where its absolute stability disappears.

This work was partially supported by Fundamental Researches State Fund of Ukraine (project \# F28/251).

\bibliography{P1}

\end{document}